\documentclass{emulateapj}

\begin{document}
\title{Can Stellar Mixing Explain the Lack of Type Ib Supernovae in Long-Duration GRBs?}

\def\msun{{\rm ~M}_{\odot}}
\def\rsun{{\rm ~R}_{\odot}}
\def\myr{{\rm ~Myr}}
\def\mdot{\dot M}
\def\mpy{{\rm ~M}_{\odot} {\rm ~yr}^{-1}}

\author{Lucille H. Frey\altaffilmark{1,2}, Chris L. Fryer\altaffilmark{3,4,5}, and Patrick A. Young\altaffilmark{6}}

\altaffiltext{1}{HPC-3, Los Alamos National Laboratory, Los Alamos, NM 87545, USA}
\altaffiltext{2}{Department of Computer Science, The University of New Mexico, Albuquerque, NM 87131, USA} 
\altaffiltext{3}{CCS-2, Los Alamos National Laboratory, 
Los Alamos, NM 87545, USA}
\altaffiltext{4}{Department of Physics and Astronomy, The University of New Mexico,
Albuquerque, NM 87131, USA} 
\altaffiltext{5}{Department of Physics, The University of Arizona,
Tucson, AZ 85721, USA} 
\altaffiltext{6}{School of Earth and Space Exploration, Arizona State University, Tempe, AZ 85276, USA}

\begin{abstract}

The discovery of supernovae associated with long-duration gamma ray 
burst observations is primary evidence that the progenitors of these 
outbursts are massive stars. One of the principle mysteries in 
understanding these progenitors has been the fact that all of these 
gamma-ray burst associated supernovae are Type Ic supernovae, with 
no evidence of helium in the stellar atmosphere. Many studies have 
focused on whether or not this helium is simply hidden from spectral 
analyses. In this Letter, we show results from recent stellar models 
using new convection algorithms based on our current understanding 
of stellar mixing. We demonstrate that enhanced convection may lead 
to severe depletion of stellar helium layers, suggesting that the 
helium is not observed simply because it is not in the star. We also 
present light-curves and spectra of these compact helium-depleted 
stars, compared to models with more conventional helium layers.

\end{abstract}


\section{Introduction}

The association of supernovae occurring simultaneously with
long-duration gamma-ray bursts (LGRBs) provides one of the strongest
pieces of evidence behind a massive star progenitor for these bursts;
see \cite{Woo06} and \cite{Hjo11} for reviews.  To ensure that the jet
is driven long enough to plow through the surrounding star, LGRB
progenitors must not have extended hydrogen envelopes.  Progenitors
of LGRBs typically invoke either strong winds, or more likely, binary
interactions, to remove the hydrogen
envelope~\citep{FWH99,Woo06,Fry07}.  The standard engine behind LGRBs
invokes the collapse of these progenitors down to a black hole, where
the accretion onto the black hole drives the LGRB jet.  These systems,
typically arising from massive stars above 20\,M$_\odot$, have large
helium shells.  Without Wolf-Rayet mass-loss from the uncovered helium
star, the helium mass for the \cite{Woo02} models predict helium
masses in the 1.0-25\,M$_\odot$ range for low-metallicity stars
between 15 and 100\,M$_\odot$ (the likely progenitor range for LGRBs and
normal Ib/c supernovae).

Currently, all LGRB-associated supernovae and hypernovae are helium
deficient, Type Ic, supernovae~\citep{Woo06,Fry07,San12}.  Explaining
the lack of Ib LGRB-associated supernova remains one of the primary
problems with our current understanding of LGRB progenitors.  A
possible clue to this progenitor problem may be related to an equally
puzzling problem in the ratio of normal Type Ic to normal Type Ib
supernovae.  \cite{Sma09} found that the number of Ic supernovae
outnumbered the Ib supernovae.  A more extensive, volume-limited,
sample found that Type Ic supernovae are twice as frequent as Ib
supernovae~\citep{Smi11}.

One solution to these progenitor problems is to fine-tune the mass-loss
from bare helium cores such that these stars lose their helium shells
in a Wolf-Rayet phase while still retaining enough mass to collapse to
a black hole.  If true, there should be a strong metallicity
dependence on the Ic/Ib rate.  An alternate solution is to hide this
helium.  \cite{Hac12} have calculated the maximum mass of helium that
can be ``hidden'' in a Type Ic supernova, placing limits between
$\sim$0.06-0.14\,M$_\odot$.  But \cite{Des12} have argued that mixing
in the supernova can alter the excitation level of the helium,
effectively hiding it.  Coupled with winds, this explosive mixing 
might be able to explain the high ratio of Ic to Ib supernovae.  
But, especially for the more massive progenitors expected in LGRBs, 
mixing in the supernova explosion is unlikely to explain the fact 
that all LGRB-associated supernovae lack helium lines.

We have discussed two possible solutions that seek to explain the lack
of helium in supernova spectra: fine-tuned winds ejecting the helium
prior to collapse and mixing in the supernova explosion making it
difficult to observe the helium lines.  We suggest a new, third solution.  
Prior to collapse, the star might efficiently burn the helium into
heavier elements.  Currently stellar structures are limited to the
structure produced by mixing length theory, a recipe designed to mimic
hydrodynamic mixing in stars.  Using three-dimensional hydrodynamics as a
guide~\citep{MA2007}, new algorithms are being developed to model
stellar mixing~\citep{You05,AMY2009}.  In this Letter, we study massive
star progenitors under these new mixing algorithms.  We find that the
latest improvements in the algorithm for mixing in stars burn much of
the helium, leaving behind a helium shell mostly comprised of
heavier elements and demonstrating the viability of this third
solution.  In Section~\ref{sec:star}, we discuss this mixing and its
effects on the helium shell layer.  We conclude with light-curve
calculations of these progenitors, demonstrating the extent to which
these new models will change our picture of both LGRB progenitors and
Ib/c supernovae in general.

\section{Stellar Mixing and the Helium Shell}
\label{sec:star}

Most stellar models have relied upon modified variants of basic mixing
length theory~\citep{Bohm58} to include the complex physics of
turbulent convection in the evolution of stars.  But recent
three-dimensional models of this convection argue that the simple mixing theory
is unable to capture the full physics behind turbulent
convection~\citep{MA2007,AMY2009,AMY2010,AM2011}.  These three-dimensional
studies have led to a revision of the theory behind turbulent
convection in stars, focusing on both connections to the Kolmogorov 
theory of turbulent cascade~\citep{AMY2009,Kol1962} and to the Lorenz 
strange attractor~\citep{AM2011,Lor1963}.

The TYCHO code has been upgraded with an algorithm based on a physical
analysis of three-dimensional hydrodynamic simulations of convection, not an
astronomically calibrated mixing-length theory. It includes
non-locality and time dependence of flow, dynamical acceleration,
turbulent dissipation, Kolmogorov heating, compositional effects and
dynamically defined boundary conditions (instead of parameterized
overshooting schemes), all in a single, self-consistent
formulation. TYCHO also includes a 177 isotope nuclear network.  
The code is regularly tested against observations of
double lined eclipsing binaries and cluster isochrones to ensure
consistency and accuracy. It produces superior fits to observational
test cases without adjustment of parameters~\citep{You05}.

With this code, we have run four stellar model progenitors for Type Ib/c
supernovae with four different zero-age main sequence masses: 15, 21, 23,
and 27\,M$_\odot$.  These stars are modeled to collapse with final masses
of 12.1, 12.8, 15.4, 16.7\,M$_\odot$ respectively.  In all cases, these
stars retain some hydrogen at collapse and, without additional mass
loss (e.g. from a binary common envelope phase or higher mass-loss
rate), these stars will produce Type II supernovae.  Many GRB progenitors 
require binary interaction and we assume this interaction leads to 
the ejection of the hydrogen envelope.  However, it is less likely that 
the common envelope will eject the helium envelope.

\begin{figure}
\epsscale{0.75}
\plotone{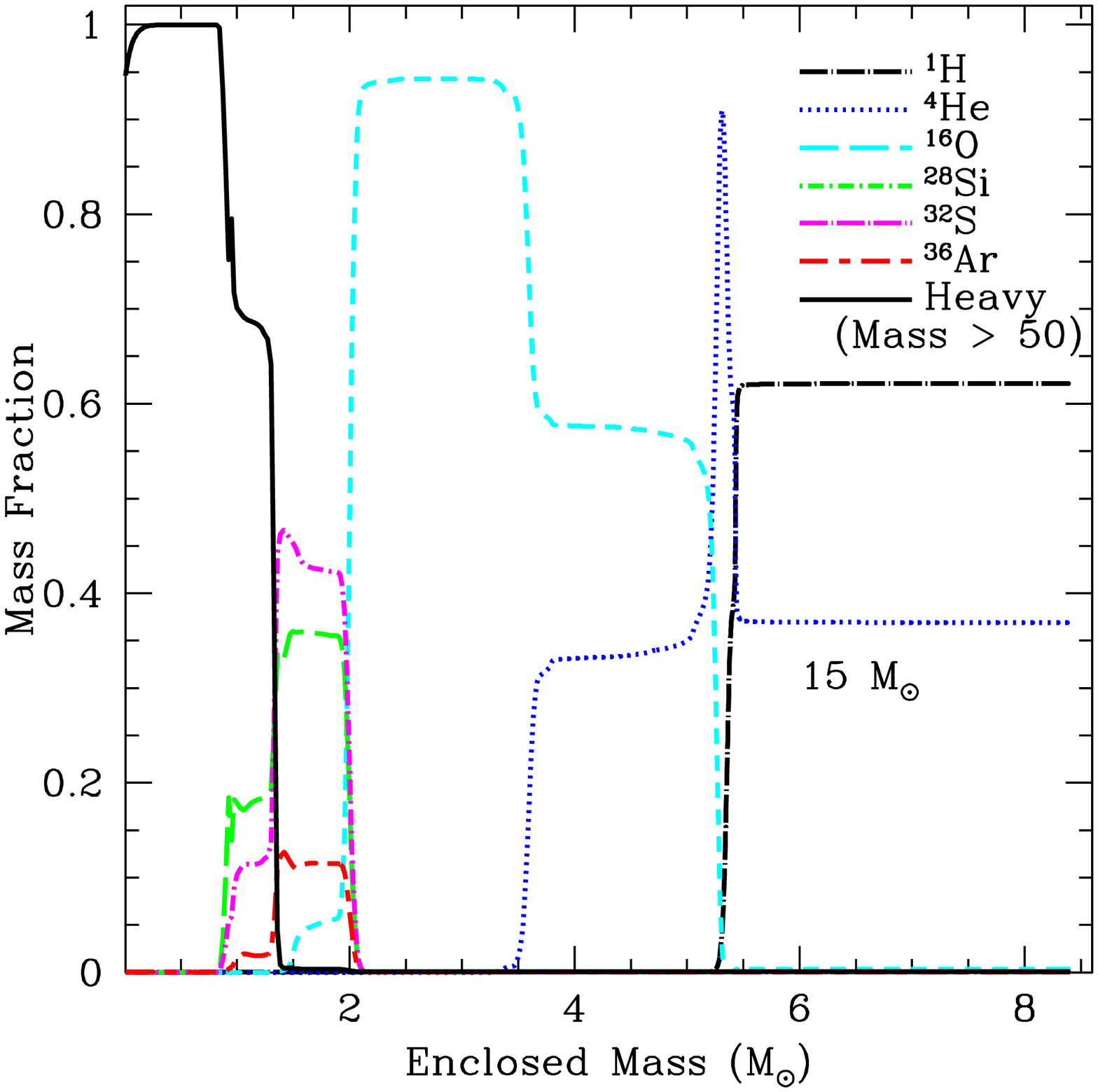}
\plotone{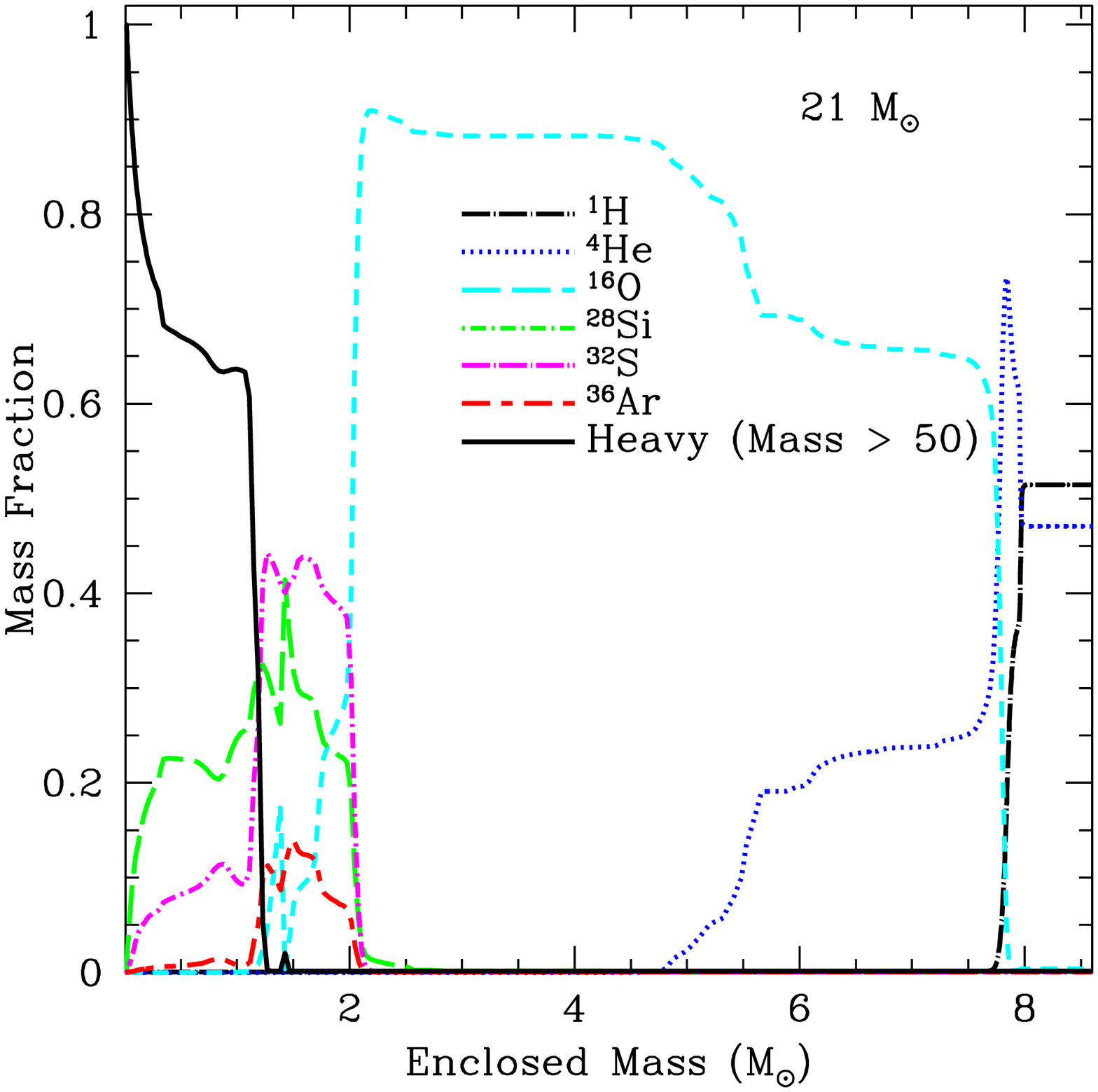}
\plotone{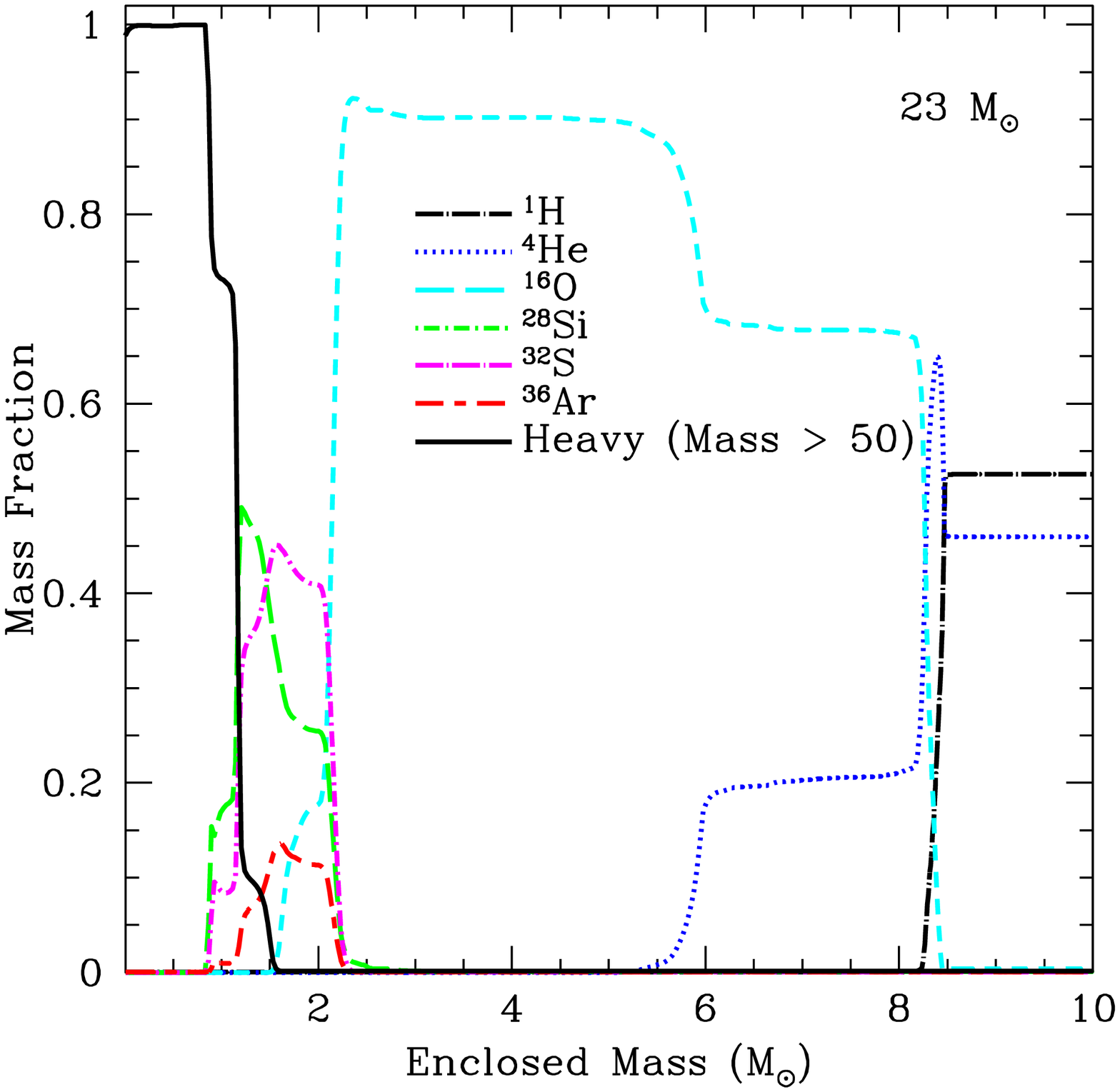}
\plotone{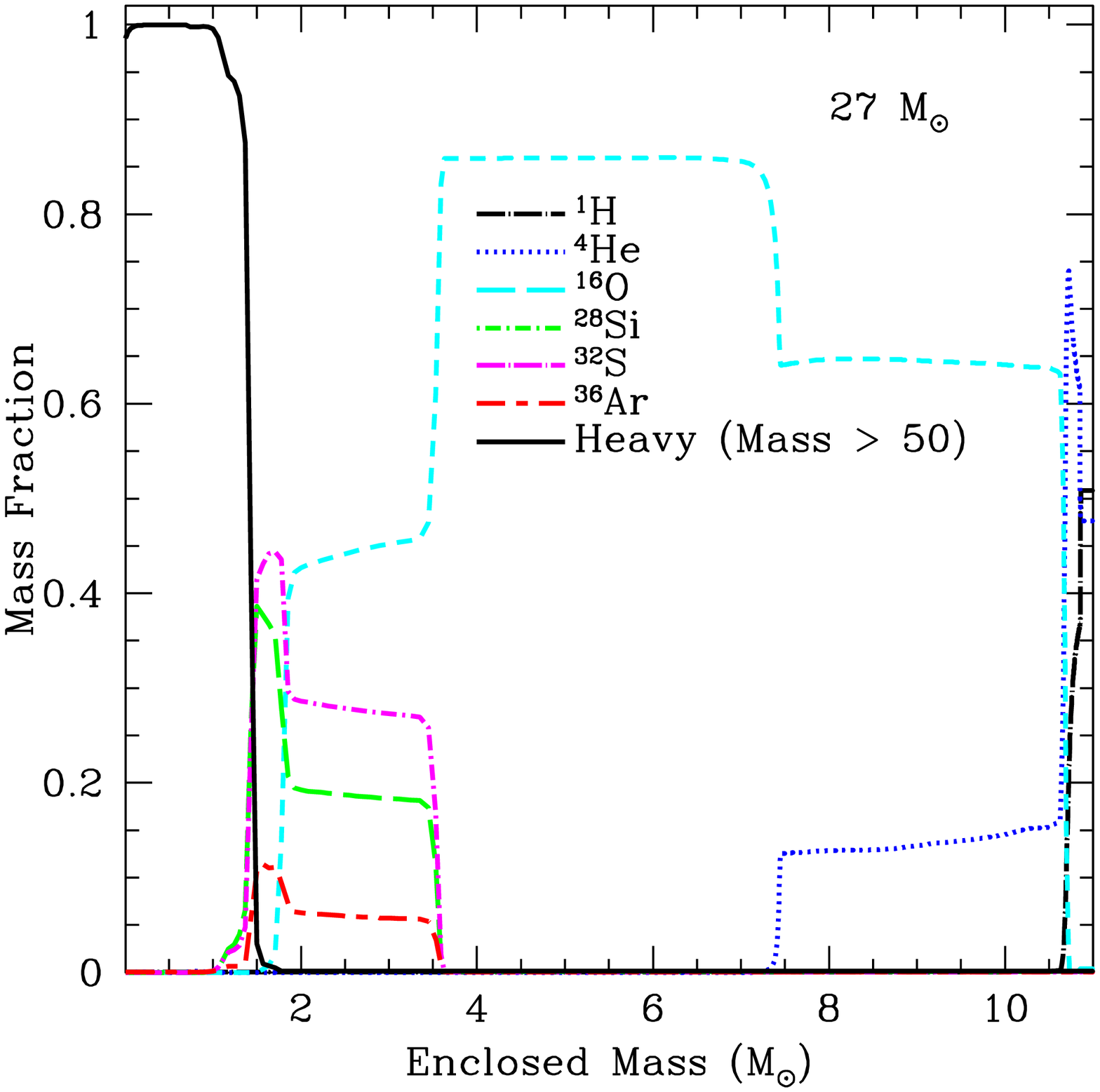}
\epsscale{1.0}
\caption{Hydrogen, helium, oxygen, silicon, sulfur, argon and heavy
  (atomic mass above 50) abundances for 15, 21, 23, and 27\,M$_\odot$
  stars with enhanced mixing.  Above 20\,M$_\odot$, the convection in
  the helium layer burns much of the helium.  This burning becomes
  more complete with more massive progenitors.}
\label{fig:abun}
\end{figure}

\begin{figure}
\epsscale{0.75}
\plotone{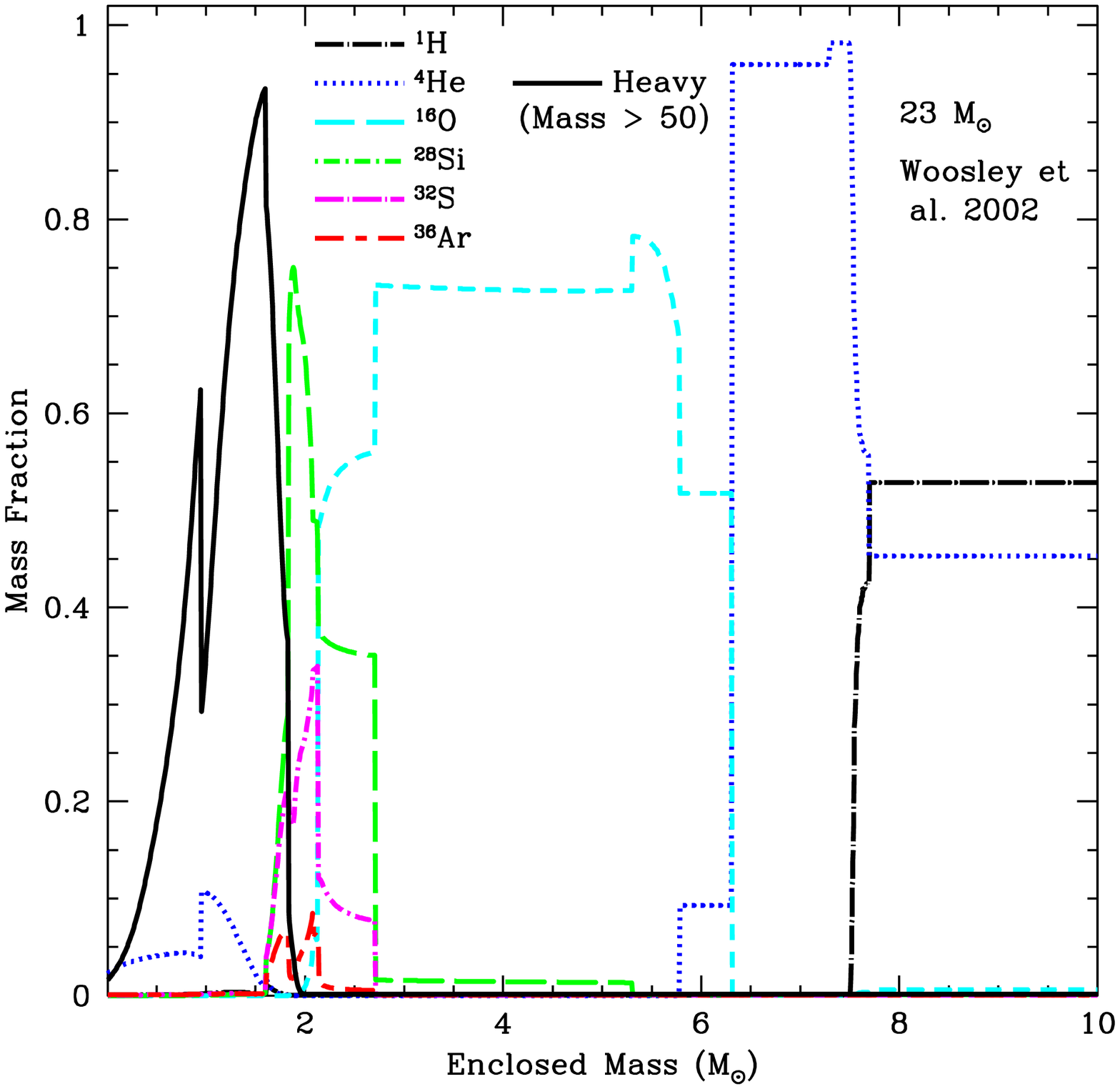}
\plotone{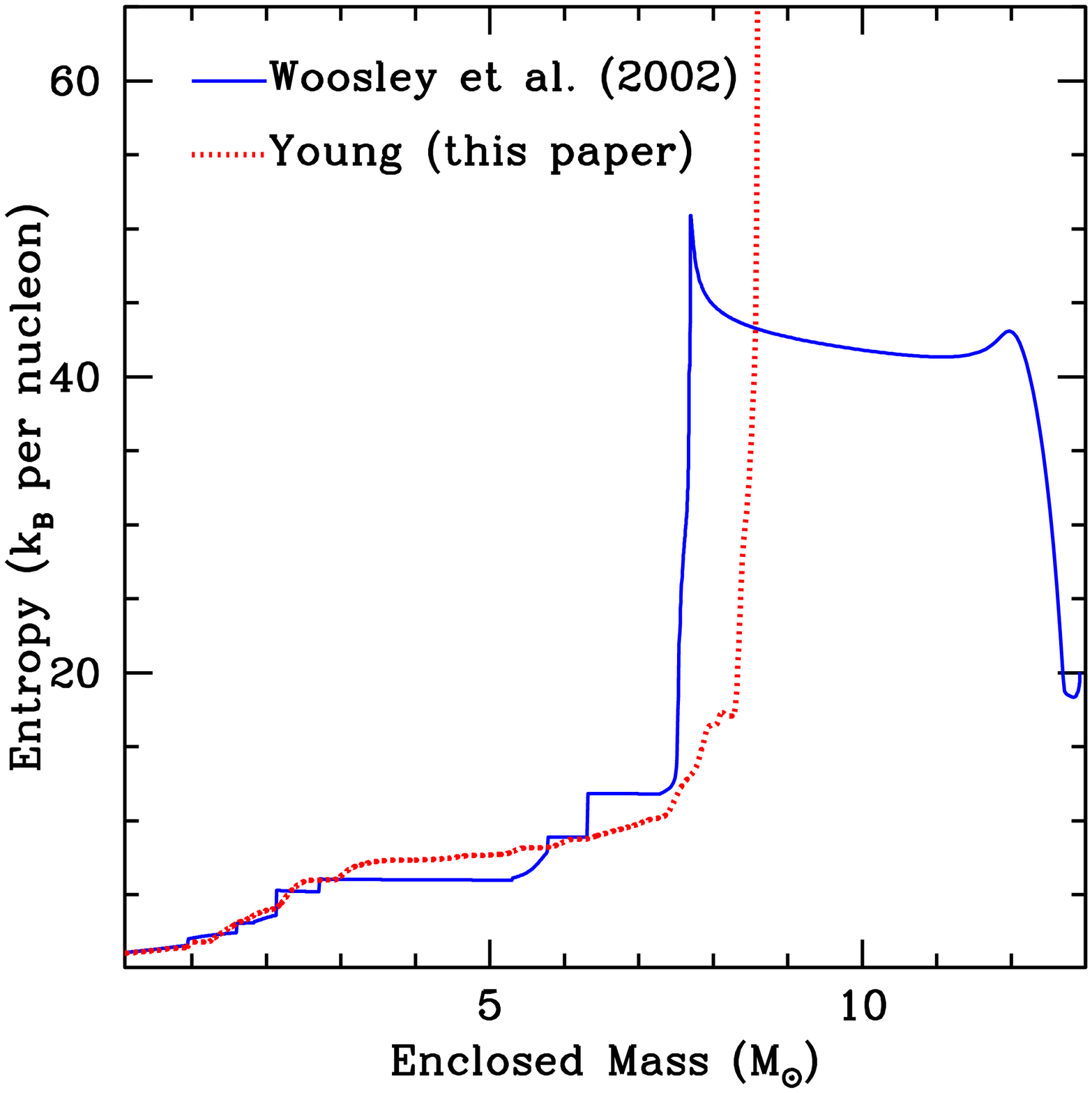}
\plotone{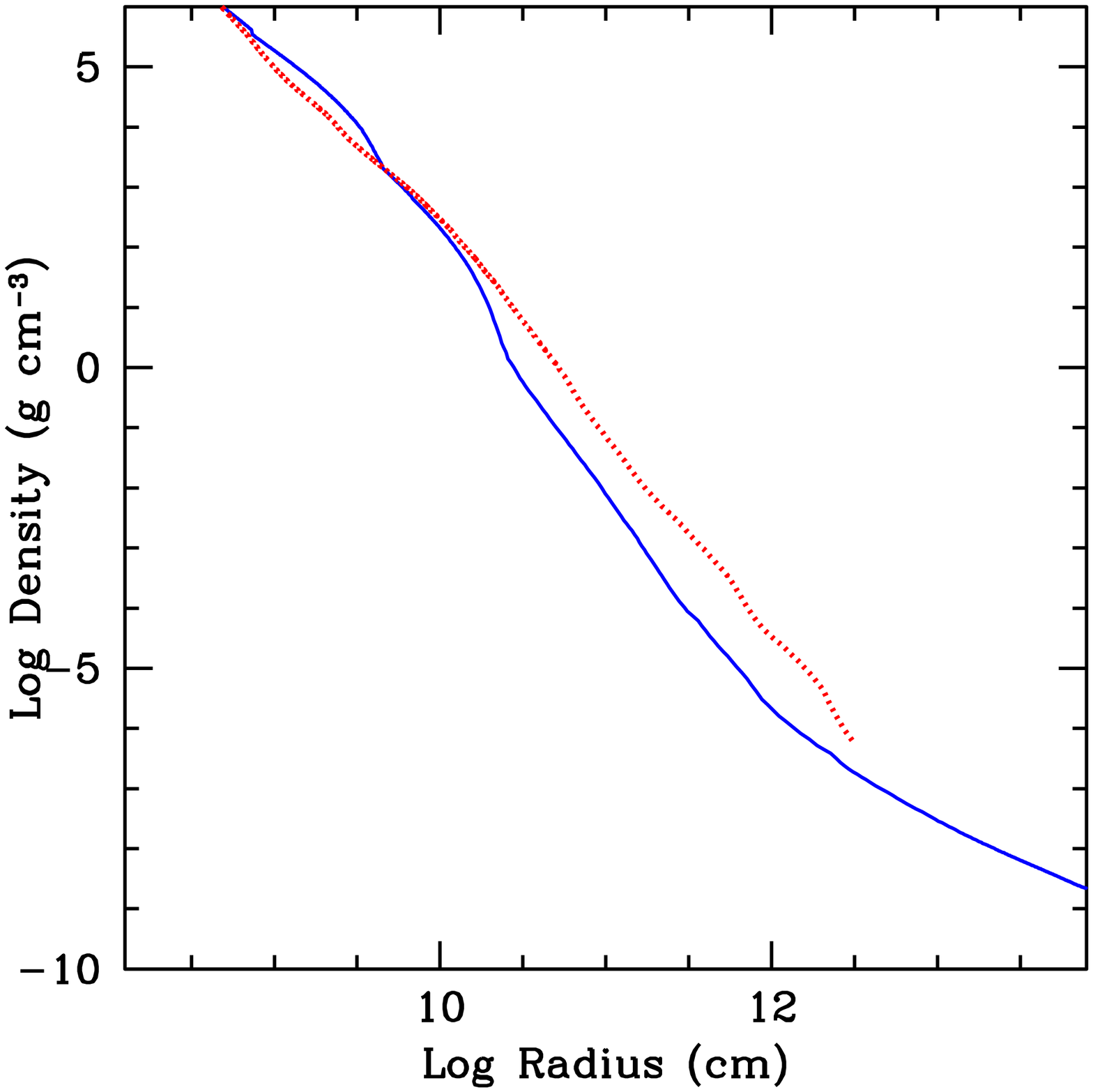}
\epsscale{1.0}
\caption{Top panel: same abundances (H,He,O,Si,S,Ar,heavies) as
  Figure~\ref{fig:abun} for a 23\,M$_\odot$ progenitor from \cite{Woo02}.  
  Note the clear helium shell.  Mixing also produces very
  different stellar structures.  The Middle panel shows the entropy
  profile.  In the ``helium-shell'' region, the entropy is lower in
  our improved convection models when compared to the 23\,M$_\odot$
  model using mixing length theory.  The corresponding density profile
  (bottom panel) is nearly an order of magnitude higher in our
  improved convection model over mixing length theory.}
\label{fig:comp}
\end{figure}

The abundance profiles (H, He, O, Si, S, Ar and heavy elements) for
our four models are shown in Figure~\ref{fig:abun}.  The 15\,M$_\odot$
star retains a normal helium shell.  But above 20\,M$_\odot$, mixing
in the helium shell burns much of the helium into oxygen.  This
burning becomes more complete with increasing progenitor mass.  By
27\,M$_\odot$, the helium fraction in the ``helium'' shell is below
15\%.  These abundances are vastly different from simulations~\citep{Woo02}
 using standard mixing length theory plus parameterized overshooting 
(Figure~\ref{fig:comp}).  Mixing length theory produces, for a
23\,M$_\odot$ star, a nearly 2\,M$_\odot$ helium layer that is over
90\% helium.  With our new model, a 2\,M$_\odot$ helium layer exists,
but it is only $\sim$20\% helium.

The mixing also changes the structure of the star.  Figure~\ref{fig:comp} 
shows the differences in the entropy and density of our 23\,M$_\odot$ and 
the \cite{Woo02} 23\,M$_\odot$.  Our star has lower entropy 
in the helium layer, producing a more compact star with densities nearly 
an order of magnitude higher in this region.  The corresponding maximum 
stellar radius is over an order of magnitude lower.

\section{Supernova Observations}
\label{sec:sne}

To illustrate the role these different abundances have on supernova
observations, we focus on the 23\,M$_\odot$ progenitor, collapsing
it with the same technique described in Young \& Fryer (2007).  The
collapse code is a one-dimensional Lagrangian code developed by Herant
et al. (1994).  This code includes three-flavor neutrino transport
using a flux-limited diffusion calculation and a coupled set of
equations of state to model the wide range of densities in the
collapse phase (see Herant et al. 1994; Fryer et al. 1999 for
details).  After the collapse, bounce, and formation of the
proto-neutron star, we stop the code and remove the neutron star.  We
then artificially inject energy into the innermost zones to drive an
explosion.  For this model, we produced a 5e51\,erg explosion.  To
study the effect of abundance differences on the light-curve and
spectra, we run two models: one is our new 23\,M$_\odot$
progenitor, the other is this same progenitor with the helium 
core abundances altered to match those from the \cite{Woo02} 
23\,M$_\odot$ progenitor (subsequently referred to as the Young and Woosley 
models, respectively).  In this manner, we can focus only on the
abundance differences and not on the structure differences we also
discussed in Section~\ref{sec:star}.

After the launch of the explosion, each model is mapped into RAGE
(Radiation Adaptive Grid Eulerian), a Eulerian radiation-hydrodynamics
code which includes adaptive mesh refinement \citep{Git08}, where it
is run out to several years.  We post-process data from individual
timesteps with the SPECTRUM code, which uses monochromatic opacities
to calculate spectra and lightcurves.  SPECTRUM allows us to study
emission and absorption as a function of radius, as well as the
effects of individual elements on observable spectra.  This pipeline
is described in detail in \cite{Frey13}.

\begin{figure}
\epsscale{0.75}
\plotone{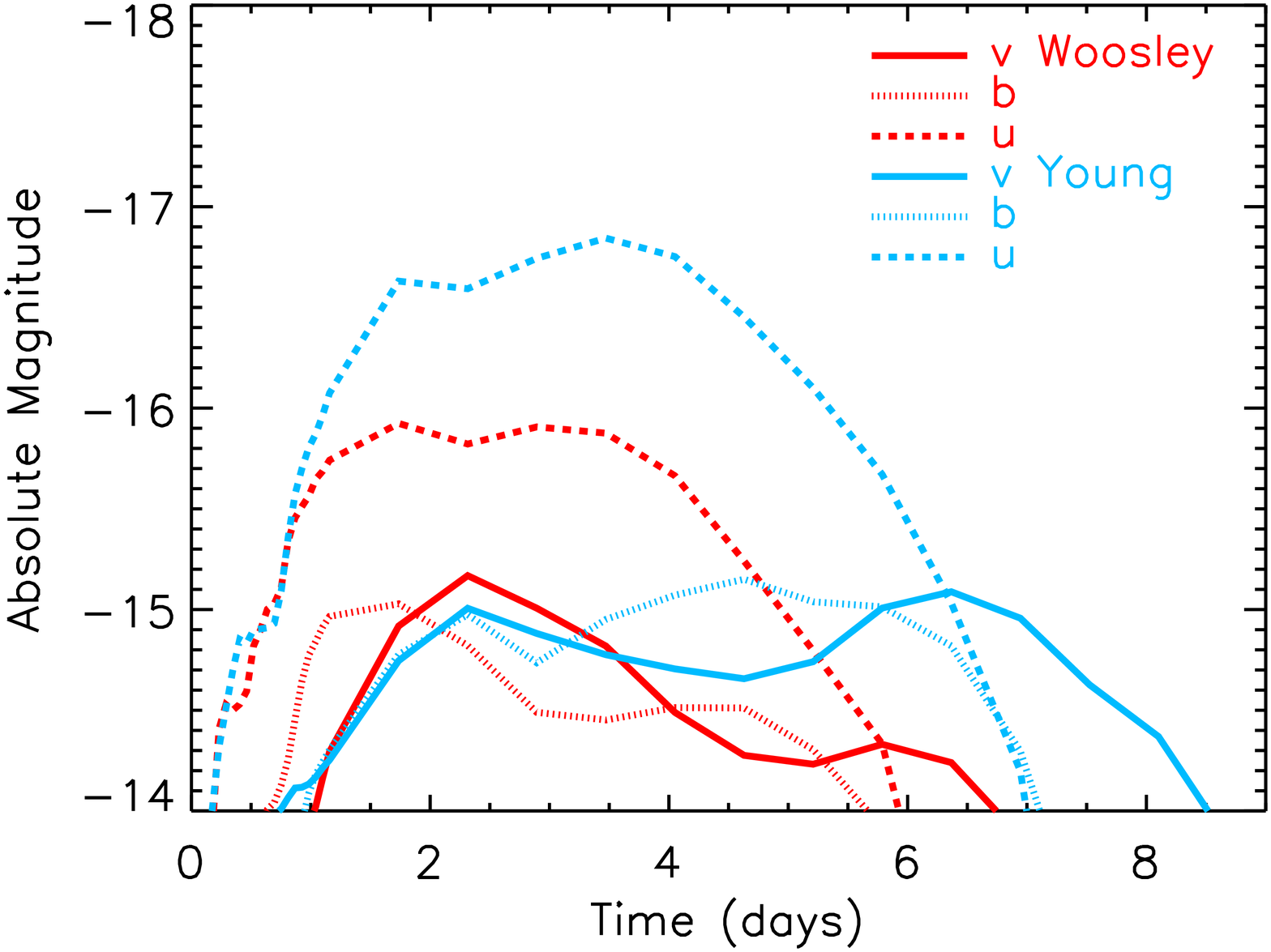}
\plotone{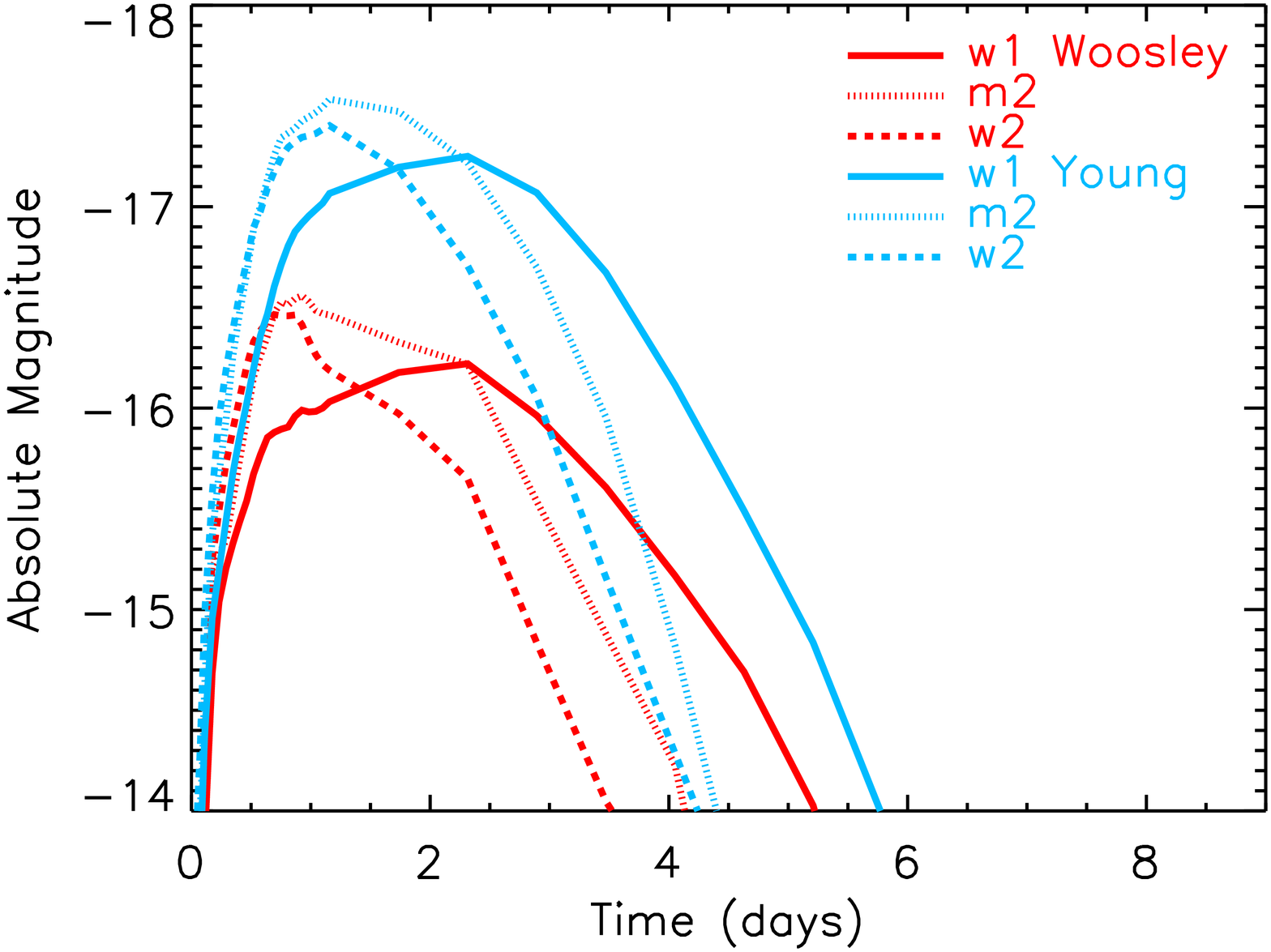}
\epsscale{1.0}
\caption{UV and optical lightcurves for the Young and Woosley models, created using the Swift filters for the v (5468 \AA), b (4392 \AA), and u (3465 \AA) bands, top, and the swuw1 (2600 \AA), swum2 (2246 \AA), and swuw2 (1928 \AA) bands, bottom. }
\label{fig:uvopt}
\end{figure}

To demonstrate the effects of the two different helium core
abundances, we compare the resulting spectra, and lightcurves in
several different bands.  The compact nature of this progenitor
produces a very short-duration burst, with the time from first peak in
the v-band to its decay by over a magnitude lasting less than 10 days
(Figure~\ref{fig:uvopt}).  We are using identical stellar structures, so
it is not surprising that the UV/optical lightcurves in
Figure~\ref{fig:uvopt} from the Young model are the same shape
as the Woosley model.  One obvious difference can be seen in the luminosities, 
as the Young model produces peak spectra at high energies (u band and higher) 
that are one magnitude dimmer than the Woosley model.  

\begin{figure}
\epsscale{0.75}
\plotone{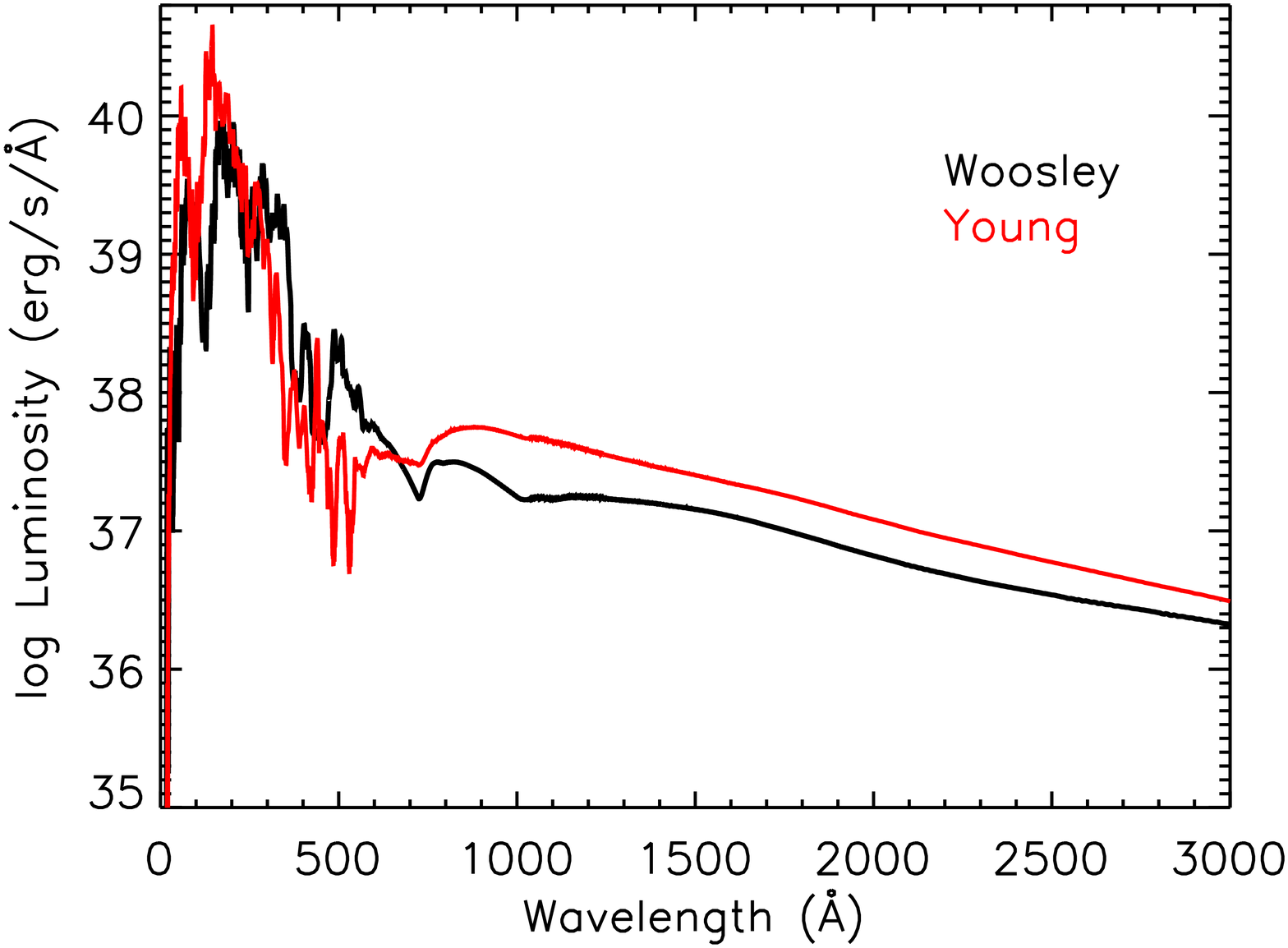}
\plotone{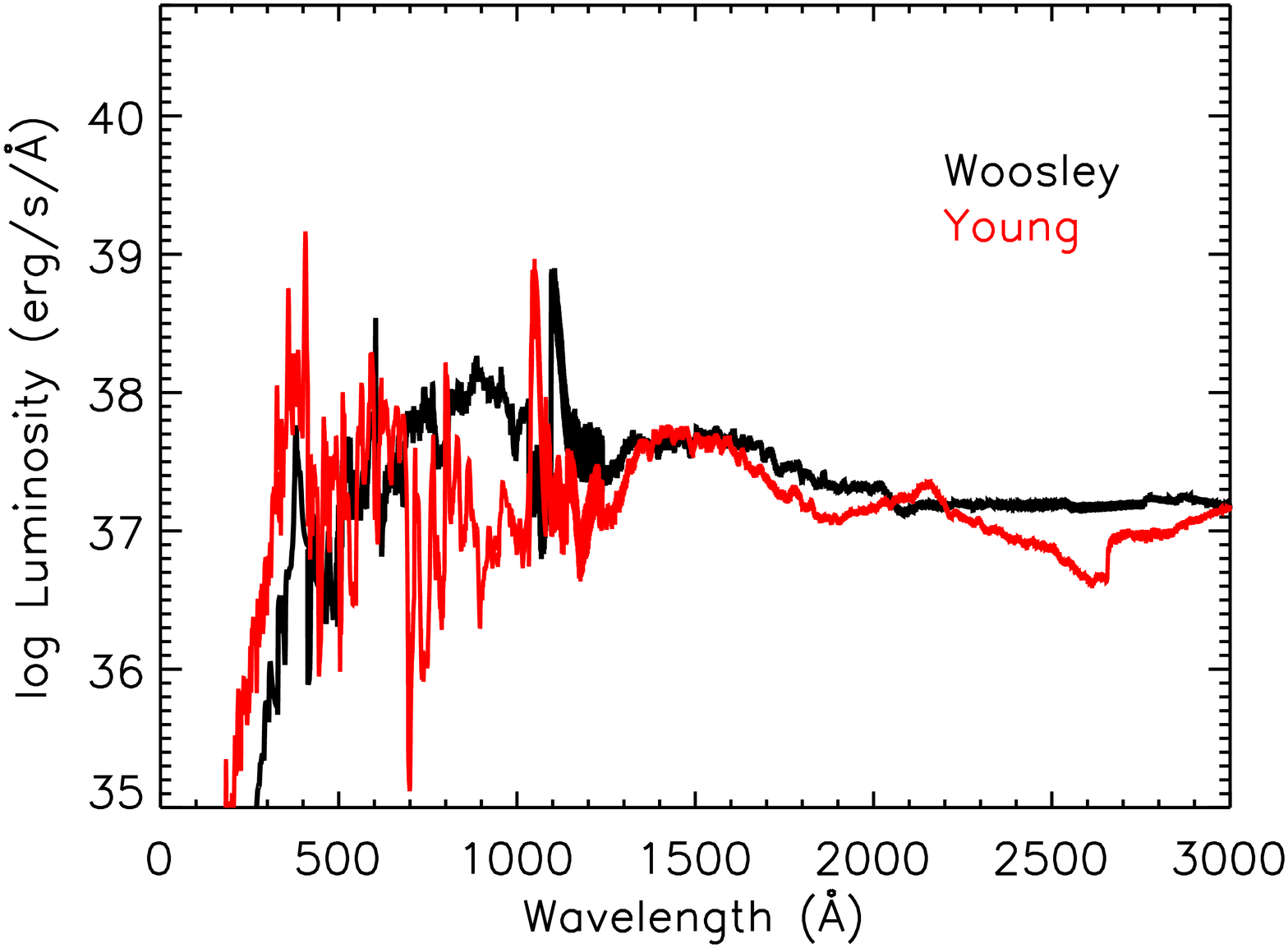}
\epsscale{1.0}
\caption{Spectra from the Young and Woosley models at 1e3 s (top) and 1e4 s (bottom).}
\label{fig:spec}
\end{figure}

The differences are more striking in the spectra.  Our SPECTRUM code calculates 
opacities by mixing opacities from individual elements, so we can artificially modify 
this algorithm to study the emission and absorption due to single elements by setting 
the opacity for a single element to zero.  This allows us to compare spectra calculated 
without helium or oxygen to complete spectra and easily identify emission and absorption 
from those elements.  As expected, the Young model shows much stronger effects from 
oxygen and weaker helium emission and absorption than the Woosley model.  At 1e3s (Figure~\ref{fig:spec}, top), we see 
helium absorption in both models around 400 \AA, but in the Young model this is combined with 
oxygen absorption in the same region.  At 1e4s (Figure~\ref{fig:spec}, bottom), the oxygen absorption around 2600 \AA  appears only 
in the Young model while a helium absorption feature at 2000 \AA  occurs in both.

\section{Conclusion}

Our new mixing algorithm produces a very different picture of the
structure and composition of Ib/c SN progenitors.  While one dimensional 
simulations cannot capture all of the complex physical processes involved in 
stellar evolution, our algorithm is based on the three-dimensional effects of convection 
and provides good fits to observational test cases~\citep{You05}.
In addition to
changing the structure of the star, this algorithm produces a ``helium
shell'' which is increasingly dominated by oxygen as the mass of the
star increases.  We expect stars with masses above $\sim$20\,M$_\odot$
which lose their hydrogen envelopes (either through common envelope
evolution or winds) to produce Ic supernovae.  The standard long-duration GRB
engine assumes that the progenitor star must collapse to a black
hole, arguing that long-duration GRBs are only produced by stars 
more massive than $\sim$20\,M$_\odot$~\citep{FWH99}.  In this manner 
our new progenitors, employing a more physically-based algorithm 
for stellar convection, provide a natural explanation for why 
GRB-associated supernovae are all Type Ic.  

Depending upon the initial mass function for stars, the minimum mass
for core-collapse and the minimum mass for black hole formation, stars
above $\sim$20\,M$_\odot$ can produce 10-40\% of all supernovae.  If
Ib/c supernovae were only produced in systems with large winds, we would 
expect these supernovae to be limited to stars above
$\sim$20\,M$_\odot$, arguing that all Ib/c supernovae would actually
be Ic supernovae.  Many Ib/c supernovae are produced through common
envelope mass ejection, producing the combination of Ib and Ic
supernovae.  Nevertheless, with our new progenitor stars, it is not
surprising that Ic supernovae dominate the Ib/c rate.

\acknowledgements We thank Wesley Even of CCS-2 at LANL for developing and 
implementing the code used to map explosion profiles from the collapse code 
into RAGE.  Work at LANL was done under the auspices
of the National Nuclear Security Administration of the U.S. Department of 
Energy under contract No. DE-AC52-06NA25396.  Young was supported in part 
by NSF Grant No. 0807567.

{}

\end{document}